\documentclass[aps,amsmath,amssymb,showpacs,showkeys,floatfix,a4,nofootinbib]{revtex4}
\usepackage{epsfig} 
\usepackage{amssymb}
\usepackage[ansinew]{inputenc}

\usepackage{soul,color}
\usepackage{graphicx}
\usepackage{dcolumn}% Align table columns on decimal point
\usepackage{bm}% bold math
\usepackage{float}
\usepackage{mathrsfs}
\usepackage{tabularx}
\usepackage{midfloat}

\newcommand{\be}{\begin{equation}}
\newcommand{\ee}{\end{equation}}
\newcommand{\bea}{\begin{eqnarray}}
\newcommand{\eea}{\end{eqnarray}}
\newcommand{\ba}{\begin{array}}
\newcommand{\ea}{\end{array}}
\newcommand{\bi}{\begin{itemize}}
\newcommand{\ei}{\end{itemize}}
\newcommand{\lan}{\langle}
\newcommand{\ran}{\rangle}
\begin{document}

\title{Heavy ion charge exchange reactions and the link with $\beta$ decay
processes}
 
\author{M.Colonna$^{1,5}$, J.I.Bellone$^{1,2,5}$, S.Burrello$^{1,5}$,
Jos\'e-Antonio Lay$^{3,1,5}$, H.Lenske$^{4,5}$}

\affiliation{$^1$INFN-LNS, I-95123 Catania, Italy\\
$^2$Dipartimento di Fisica e Astronomia, Universit\'a degli studi di Catania, Italy\\
$^3$Departamento de FAMN, Universidad de Sevilla, Apartado 1065, E-41080 Sevilla, Spain\\
$^4$Institut f\"{u}r Theoretische Physik, Justus-Liebig-Universit\"{a}t Giessen, D-35392 Giessen, Germany\\
$^5$NUMEN collaboration, LNS Catania}

%\maketitle % this produces the title block

\begin{abstract}
Within the DWBA framework, we develop a theoretical description of single and double heavy ion charge exchange (CE) reactions.
We show that absorption effects are particularly important for 
heavy ion reactions, leading to a noticeable reduction of the CE 
cross sections. At low momentum transfer, the single CE cross section can be
factorised, thus allowing to evaluate corresponding distortion factors and
access $\beta$ decay strengths. Applications are shown for a system of
experimental interest. Preliminary results are discussed also for double 
CE reactions, modeled as a two-step mechanims, i.e. a sequence of two
charge-changing processes.  
\end{abstract}
\maketitle % this produces the title block

\section{Introduction}
 
Nuclear charge exchange reactions offer the possibility to explore the nuclear interaction in the
spin-isospin channel, being related to excitations inducing isospin flip 
(with possibly also
spin flip), such as the Gamow-Teller resonance ($GTR$). 
%by the pioneering experiments at IUCF  \cite{Goodman:1980} initiated widespread experimental and theoretical research activities, continuing with even increasing intensity until today. 
Over the years, a wealth of data has been accumulated as reviewed e.g. in \cite{Ichimura:2006mq,Thies:2012xg,Frekers:2013yea,Frekers:2015wga,Fuji:2011}. Beyond using nucleonic probes, light ion reactions as e.g. $(^3He,{}^3H)$ have become another workhorse of the field, now reaching accuracies allowing to investigate subtle details of spectral distributions in both the $\tau_+$ and the $\tau_{-}$ branches. Soon after the first light ion studies, also heavy ions were used in charge exchange studies, as in \cite{Brendel:1988,Berat:1989amx}. 
%The link between charge-exchange resonances, beta decay and cross sections measured in charge exchange reactions has been widely explored in the past, but only in the case of single beta decay and
%single CE reactions with light projectiles. Indeed, GT and beta decay transitions are intimately related.

%It was recognized that peripheral heavy ion collisions, leading to direct reactions, are as useful for spectral studies as light ion scattering. An especially appealing aspects is the broad range of projectile-target combinations which, for example, allow to project out selectively specific features, e.g. spin flip and non-spin flip transitions \cite{Lenske:1989zz}. Nuclear spin-dynamics and the population of continuum states were central aspects of the $(^7Be,{}^7Li)$ reactions considered in \cite{Cappuzzello:2004vka,Cappuzzello:201lyi}. 
While the past experiments have been focused on single charge exchange (SCE) reactions, new territory was entered by the pilot experiment of Cappuzzello et al. \cite{Cappuzzello:2015ixp}, studying for the first time a nuclear double charge exchange (DCE) reaction. The reaction $^{18}O+{}^{40}Ca\to{} ^{18}Ne+{}^{40}Ar$ gave strong evidence for a direct reaction mechanism even for double charge exchange processes. Quite recently, the NUMEN project at LNS Catania was initiated, dedicated to investigations of SCE and DCE heavy ion reactions, elucidating and optimizing their potential 
for spectroscopic studies \cite{Cappuzzello:2018wek}, also in the perspective of probing the single/double $\beta$-like 
nuclear response. DCE experiments have been recently performed also 
with different goals: search for exotic systems (such as the tetra-neutron (4n) system in $^4He(^8He,^8Be)4n$ reactions); search for the Double GT resonance for quantitative information about the corresponding
sum-rule (for example in ($^{12}C$,$^{12}Be$) reactions) \cite{Kisa,Taka}. 

%probe of the double beta -like nuclear response (with (20Ne,20O) reactions) [27-29].
A thorough theoretical investigation of the dynamics of heavy-ion CE reactions, concerning in particular the possibility of extracting the nuclear structure information out of the total reaction cross section, is
still missing, though some progress has been made over the past year \cite{noi}. This is especially important for DCE reactions, that necessarily involve heavier projectiles than the light ones traditionally employed
in CE reactions.
The possibility to single out the relevant structure information is an essential point if one wishes to exploit the measured cross sections as stringent benchmarks to constrain the theoretical structure models and the
predicted transition matrix elements.   This would allow to shed light on yet unknown aspects of charge-exchange transitions and of the
underlying nuclear effective interaction.  Moreover, owing to the analogies between strong and weak charge exchange processes, 
DCE studies can provide useful information to improve the accuracy of the calculations 
of the nuclear matrix elements responsible for neutrinoless
double-beta decay, a quite hot subject of investigation nowadays. 

In this contribution, we briefly review some aspects of the theory for heavy ion charge exchange reactions. In particular, we discuss the possibility to single out the structure information from the reaction cross section.  Illustrative results are shown for systems recently investigated by the NUMEN collaboration.

\section{Theory of Heavy Ion Charge Exchange Reactions}\label{sec:Reaction}

Charge changing reactions 
%by strong interactions are off-shell processes mediated by the exchange of virtual particles. They 
require two reaction partners, which are acting mutually as the source or sink, respectively, of the charge-changing virtual meson fields. 
%For experimental reasons, the projectile-like ejectile should be preferentially in a particle-stable state (see, however ($d$,$^2He$) reactions \cite{Bugg:1987zr}), thus simplifying the detection. If the ejectile has only a single bound state below the particle emission threshold, the calculations and the interpretation of the spectroscopic data are especially simple.
Let us start considering ion-ion SCE reactions according to
\be\label{eq:reaction}
^a_za+{}^A_ZA\to{}^a_{z\pm 1}b+{}^A_{Z\mp 1}B
\ee
which change the charge partition by a balanced redistribution of protons and neutrons.

The differential SCE cross section for a reaction connecting the channels $\alpha$ and $\beta$ is defined as
%\bea\label{eq:xsec_gen}
%&&d2\sigma_{\alpha\beta}=\frac{m_\alpha m_\beta}{(2\pi\hbar^2)^2}\frac{k_\beta}{k_\alpha}\frac{1}{(2J_a+1)(2J_A+1)} \times
%\nonumber \\
%&&\sum_{M_a,M_A\in \alpha;M_b,M_B\in \beta}{\left|{M_{\alpha\beta}(\mathbf{k}_\alpha,\mathbf{k}_\beta)}\right|^2}d\Omega,
%\eea
\be\label{eq:xsec_gen}
d^2\sigma_{\alpha\beta}=\frac{m_\alpha m_\beta}{(2\pi\hbar^2)^2}\frac{k_\beta}{k_\alpha}\frac{1}{(2J_a+1)(2J_A+1)} \times
%\nonumber \\
\sum_{M_a,M_A\in \alpha;M_b,M_B\in \beta}{\left|{M_{\alpha\beta}(\mathbf{k}_\alpha,\mathbf{k}_\beta)}\right|^2}d\Omega,
\ee
where  $\mathbf{k}_\alpha$ ($\mathbf{k}_\beta$) and $m_\alpha$ ($m_\beta$) denote
the relative 3-momentum and reduced mass in the entrance (exit) channel $\alpha=\left\{ a,A \right\}$ ($\beta=\left\{ b,B \right\}$).
$\{J_aM_a,J_AM_A\cdots\}$ and $\{J_bM_b,J_BM_B\cdots\}$ account for the full set of (intrinsic) quantum numbers specifying the initial and final channel
states, respectively.

In distorted wave approximation, the  direct charge exchange reaction amplitude is given by the expression
\be
M_{\alpha\beta}(\mathbf{k}_\beta,\mathbf{k}_\alpha)=\lan\chi^{(-)}_\beta, bB|T_{NN}|aA,\chi^{(+)}_\alpha\ran.
\ee
The distorted waves, denoted  by $\chi^{(\pm)}_{\alpha,\beta}$ for asymptotically outgoing and incoming spherical waves, respectively, depend on the respective channel momenta $\mathbf{k}_{\alpha,\beta}$ and the optical potentials, thus accounting for initial state and final state interactions.

The charge-changing process is described by the nucleon-nucleon (NN) T-matrix $T_{NN}$. 
%Anti-symmetrization between target and projectile nucleons is taken care of by the standard procedure of attaching the spin and isospin exchange operators to the T-matrix and treating space-exchange in local momentum approximation, see e.g. \cite{Satchler:1983,Franey:1985ye,Hofmann:1998}. 
The anti-symmetrized T-matrix is given in non-relativistic momentum representation by
%\be
%\begin{split}
%&T_{NN}(\mathbf{p})=\sum_{S,T}  \big\{ V^{(C)}_{ST}(p^2)O_{ST}(1)\cdot O_{ST}(2)\\%\nonumber \\
%&+\delta_{S1}V^{(Tn)}_T(p^2)\sqrt{\frac{24\pi}{5}}Y^*_{2} (\mathbf{\hat{p}})\cdot \left[O_{ST}(1)\otimes O_{ST}(2)\right]_{2} \big\},
%\end{split}
%\ee
\be
%\begin{split}
T_{NN}(\mathbf{p})=\sum_{S,T}  \big\{ V^{(C)}_{ST}(p^2)O_{ST}(1)\cdot O_{ST}(2)%\nonumber \\
+\delta_{S1}V^{(Tn)}_T(p^2)\sqrt{\frac{24\pi}{5}}Y^*_{2} (\mathbf{\hat{p}})\cdot \left[O_{ST}(1)\otimes O_{ST}(2)\right]_{2} \big\},
%\end{split}
\label{tmatrix}
\ee
including isoscalar and isovector central spin-independent ($S=0$) and spin-dependent ($S=1$) interactions with form factors $V^{(C)}_{ST}(p^2)$, respectively, and rank-2 tensor interactions with form factors $V^{(Tn)}_{T}(p^2)$.
$O_{ST}$ denotes the spin-isospin operators
%\be\label{eq:OST}
$O_{ST}(i)=\left(\bm{\sigma}_i\right)^S\left(\bm{\tau}_i\right)^T.$
%\ee
%which describe the operator structure of both the central and tensor interactions.

In Eq.(\ref{tmatrix}) scalar products are indicated as a dot-product and the rank-2 tensorial coupling affects only the spin degrees of freedom. The subset of isovector operators, corresponding to  Fermi-type $S=0,T=1$ and Gamow-Teller-type $S=1$, $T=1$ operators contributes to the charge-changing reaction amplitudes.

The matrix element of a single charge exchange reaction, Eq.(3), can be written in slightly different form as:
\be\label{eq:DWBA_SCX}
M_{\alpha\beta}(\mathbf{k}_\alpha,\mathbf{k}_\beta)=\lan \chi^{(-)}_\beta|\mathcal{U}_{\alpha\beta}|\chi^{(+)}_\alpha\ran .
\ee
Then the nuclear structure information on multipolarities, transition strength and interactions are contained in the (anti-symmetrized) transition potential
$\mathcal{U}_{\alpha\beta}$,
%\be\label{eq:Uab_SCX}
%\mathcal{U}_{\alpha\beta}(\mathbf{r}_\alpha,\mathbf{r}_\beta)=
%\lan J_bM_bJ_BM_B|T^{(C)}_{NN}+T^{(Tn)}_{NN}...|J_aM_aJ_AM_A\ran
%\ee
depending on the channel coordinates $\mathbf{r}_{\alpha,\beta}$.

In the momentum representation, the full reaction amplitude can be rewritten as:
\be\label{eq:ME_SCE}
M_{\alpha\beta}(\mathbf{k}_\alpha,\mathbf{k}_\beta)=\int{d^3p \mathcal{U}_{\alpha\beta}(\mathbf{p})N_{\alpha\beta}(\mathbf{k}_\alpha,\mathbf{k}_\beta,\mathbf{p})  },
\ee
where the distortion coefficient
\be\label{eq:Nab}
N_{\alpha\beta}(\mathbf{k}_\alpha,\mathbf{k}_\beta,\mathbf{p})=\frac{1}{(2\pi)^3}\lan \chi^{(-)}_\beta|e^{-i\mathbf{p}\cdot \mathbf{r}}|\chi^{(+)}_\alpha\ran ,
\ee
has been introduced, showing the dressing of the nuclear transition potential by initial and final state ion-ion interactions.

The reaction kernel is given by a product of form factors:
\be\begin{split}
&U^{(ST)}_{\alpha\beta}(\mathbf{p})=
(4\pi)^2( V^{(C)}_{ST}(p^2)F^{(ab)\dag}_{ST}(\mathbf{p})\cdot F^{(AB)}_{ST}(\mathbf{p})\\%\nonumber \\
&+\delta_{S1}\sqrt{\frac{24\pi}{5}}V^{(Tn)}_{ST}(p^2)%\nonumber \\
Y^*_{2}(\mathbf{\hat{p}})\cdot \left[F^{(ab)\dag}_{ST}(\mathbf{p})\otimes F^{(AB)}_{ST}(\mathbf{p})\right]_{2}),
\end{split}
\label{eq:kern}
\ee
where $F_{ST}({\bf p})$ represents the Fourier transform of the (projectile or target) transition density.

A possible extension of the formalism to DCE reactions is based on second order perturbation theory, thus double charge exchange is depicted as a two-step process, i.e. a sequence of two single charge exchange transitions. This implies a summation 
over (virtual) intermediate states, and the treatment of the
relative distortion effects in the intermediate channel.  

\begin{figure}
\centering\includegraphics[width=.5\linewidth]{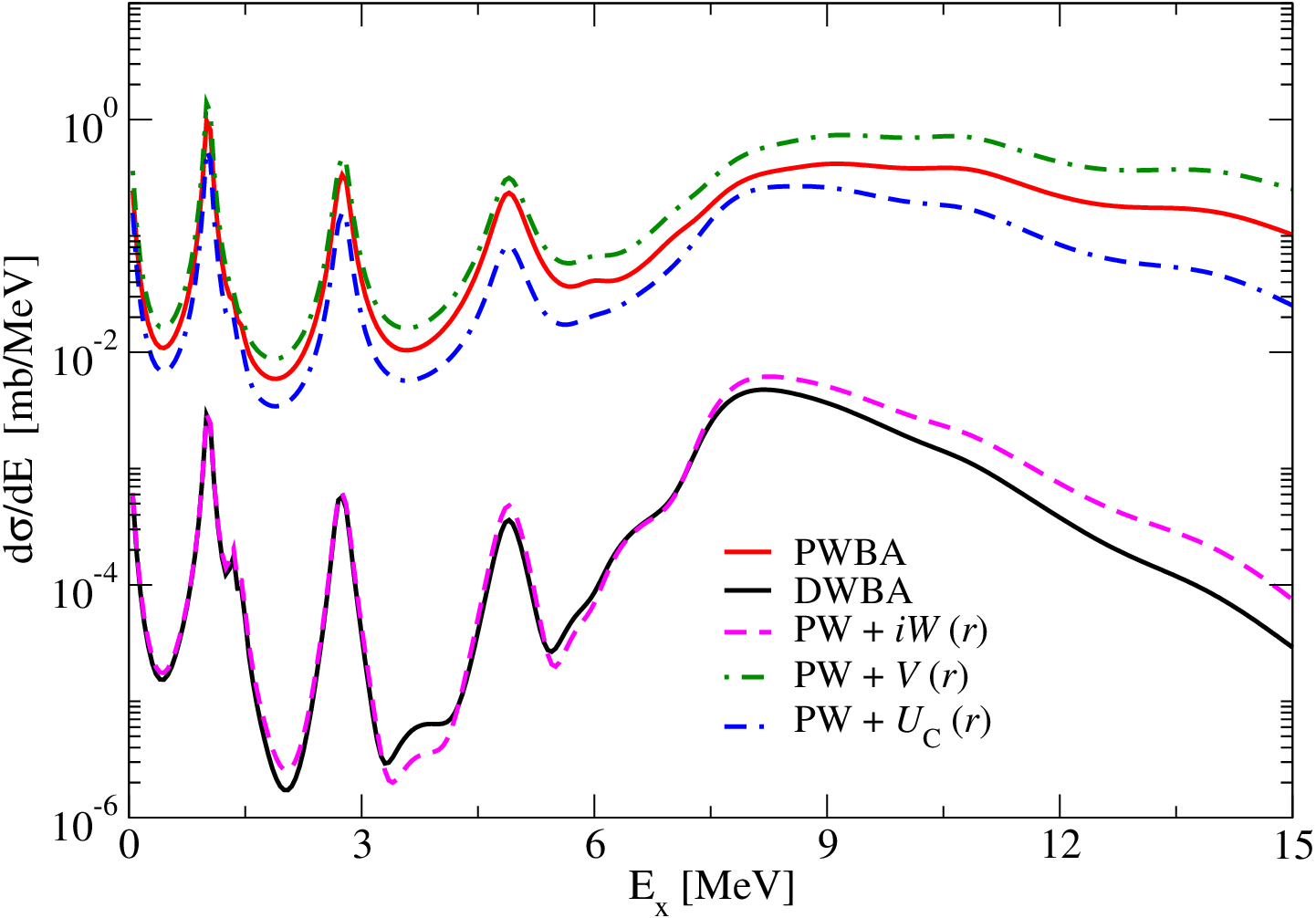}
%fig12_SCE
\caption{(Color online) Cross sections as a function of the target excitation energy, $E_x$,
for the $J^P = 1^+$ transition,
for the SCE reaction $^{40}Ca\left(^{18}O,{}^{18}F\right){}^{40}K$ reaction at $T_{lab}=270\,MeV$,
integrated over the full angular range. The different curves show the effect of Coulomb potential ($U_C(r)$), of real ($V(r)$) and imaginary ($W(r)$) components of the optical potential and of the full potential (DWBA), with respect to PWBA calculations.}
%\label{fig:myfig}
\end{figure}

\section{Results for single CE} 

As stressed above, our guiding principle to describe charge exchange reactions is direct nuclear reaction theory, based on the Distorted Wave Born Approximation (DWBA).
 Initial and final state ion-ion interactions are described by optical potentials. Microscopic optical potentials are used, obtained in the impulse approximation, 
by folding projectile and target Hartree-Fock-Bogoliubov (HFB) ground state densities with free space nucleon-nucleon 
T-matrices. 
QRPA calculations are performed for nuclear SCE transition densities and response functions, employing a G-Matrix interaction (see Ref.\cite{noi} and Refs. therein). For the sake of consistency, the same iteraction
is considered for the evaluation of the reaction kernel, Eq.(8).
More details can be found in Ref.\cite{noi}. 

As a case of physical interest, we consider throughout the SCE reaction $^{18}O+{}^{40}Ca\to ^{18}F+{}^{40}K$ at $T_{lab}=15$~AMeV.  In particular, we consider transitions leading to the  
$^{18}F$ g.s., that is a $1^+$ state, and to $J^P = 1^+$ states 
for the target.
Fig.1 represents the SCE cross section, integrated over the full angular range,  as a function of the target excitation energy, as obtained 
with the HIDEX code \cite{Cappuzzello:2004afa}.  One can identify several peaks, associated with corresponding excited states.  For comparison, calculations are performed also in the Born approximation (PWBA), and isolating the effect of the different parts of the optical potential.  One can notice that the DWBA cross section is quite suppressed with respect to PWBA results, pointing to 
strong absorption effects.  Indeed results very close to the full DWBA calculations are obtained by simply considering the imaginary part of the optical potential. We notice that strong absorption effects are peculiar of heavy ion reactions.
\begin{figure}
\centering\includegraphics[width=.5\linewidth]{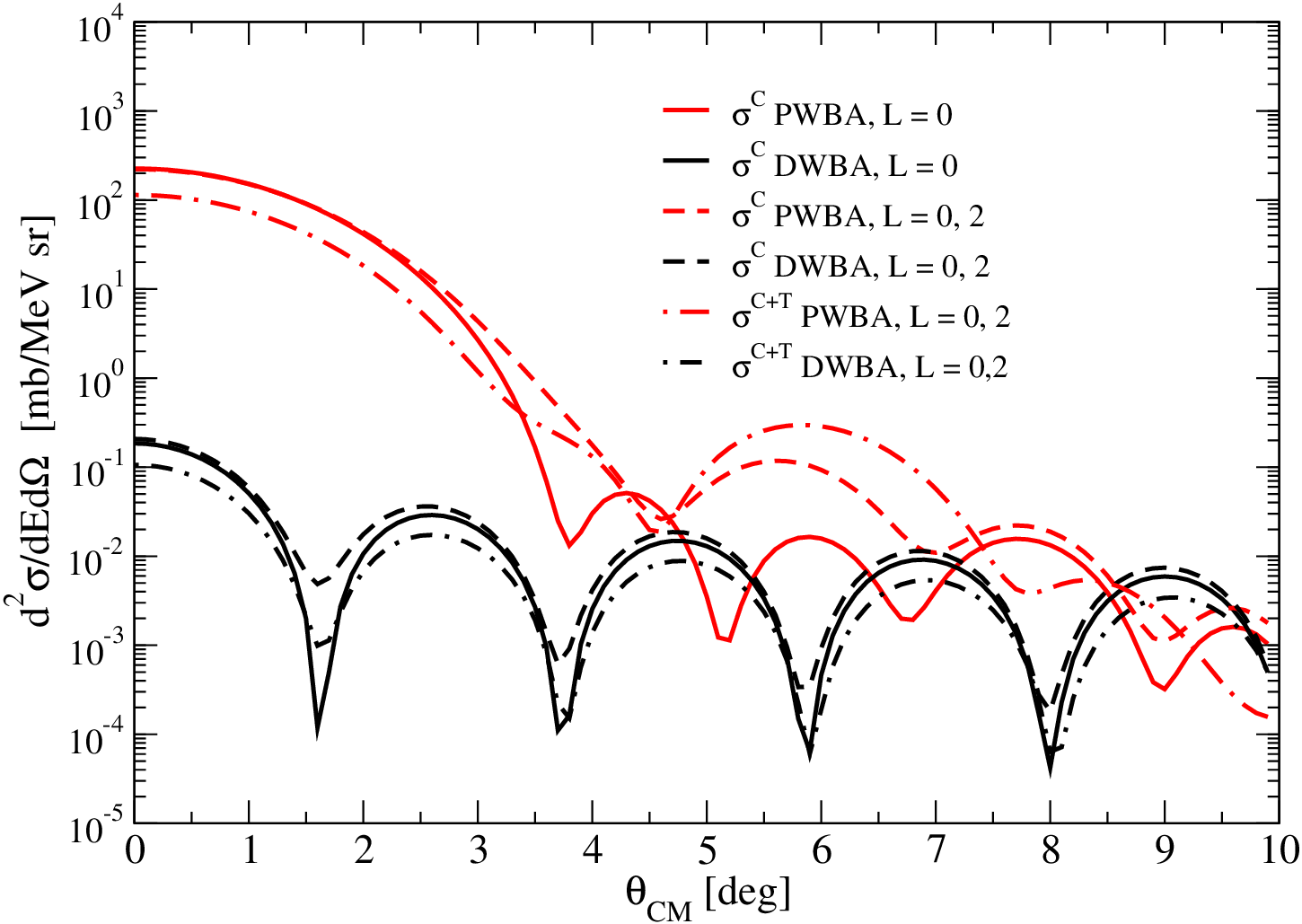}
%fig15_SCE
\caption{(Color online) Angular distribution for the target state at $E_x$ = 0.
The plot shows the effects related to central and tensor components of the nuclear interaction, for the two multipolarities allowed by $J^{\pi}=1^+$ transitions: $L=0,2$.  The system is the same as in the previous figure.}
\end{figure}  
Fig.2 displays the angular distribution obtained for the $1^+$ target state at the lowest excitation energy. The contribution of the two allowed multipolarities, as well as the effect of central and tensor components of the nuclear interaction, are shown on the figure. 
Also in this case PWBA and DWBA results are compared, allowing one to appreciate how the diffraction pattern is affected by the distortion effects. One can also observe that at small angles L = 0 transitions dominate and that the effects of the tensor component of the nuclear interaction are rather small.

\section{Cross section factorization} 
Explicit simplified expressions can be derived for forward angle cross sections, owing to the small momentum transfer ${\bf q}_{\alpha\beta}$. 
Indeed, in this case the reaction kernel in Eq.(8) can be factorized into the product of its on-shell value, ${\mathcal U}({\bf q}_{\alpha\beta})$, and a coefficient, $h({\bf q})$,  
depending on the off-shell momentum  ${\bf q} = {\bf p} - {\bf q}_{\alpha\beta}$. Then, after the integration is performed, Eq.(6) 
can be written as $M_{\alpha\beta} = {\mathcal U}({\bf q}_{\alpha\beta})
(1-n_{\alpha\beta})$, leading to the distortion factor $f_{\alpha\beta} = 
 |1-n_{\alpha\beta}|^2$ in the cross section.

Analytical calculations can be performed in the black disk (BD) approximation and considering a Gaussian fit of the transition
form factors.   Results for the distortion factor are displayed in 
Fig.3 (left panel), as a function 
of the BD radius $R_{abs}$.  For the reaction considered in our study, 
$R_{abs} \approx 8$  fm, as it can be extracted from the absorption
cross section. Correspondingly, the suppression
factor is found to be   8.14 $~10^{-4}$, 
in good agreement with the numerical DWBA/PWBA
result, $f_{BD}(numerical)|_{Ex=0} = 8.35~10^{-4}$, as it can be
extracted from the ratio between DWBA and PWBA calculations
at zero angle (see Fig.2).  Fig.3 (right panel) shows the distortion factor obtained for the reaction considered so far and for the system
$^{18}O+{}^{116}Sn$, as a function of the beam energy.
Larger distortion effects are observed for the heavier system. 
\begin{figure}
\begin{center}
\vskip0.5cm
%\centering\includegraphics[width=.3\linewidth]{fig18_SCE_fin}
\includegraphics[width=.4\linewidth]{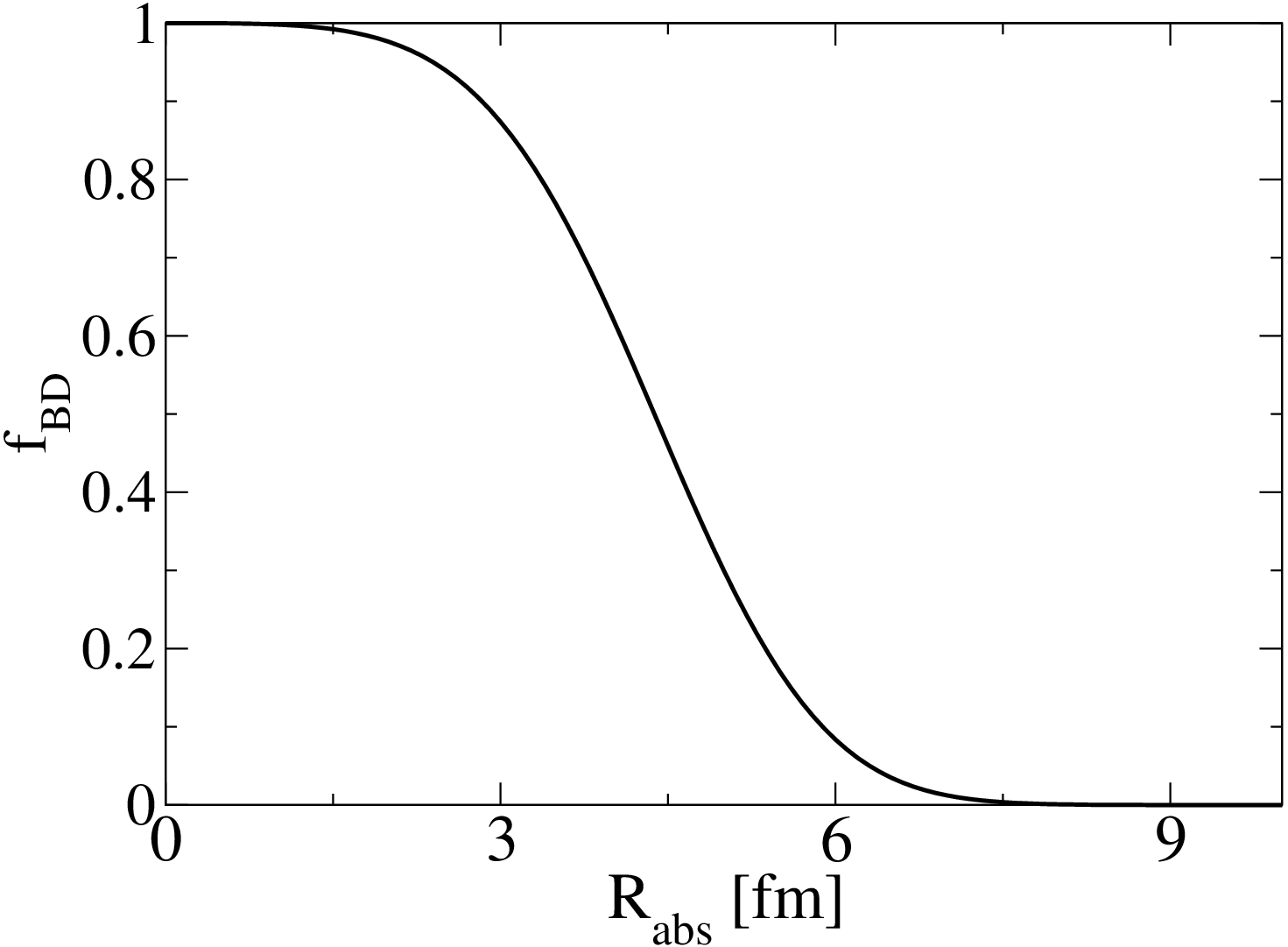}
\includegraphics[width=.43\linewidth]{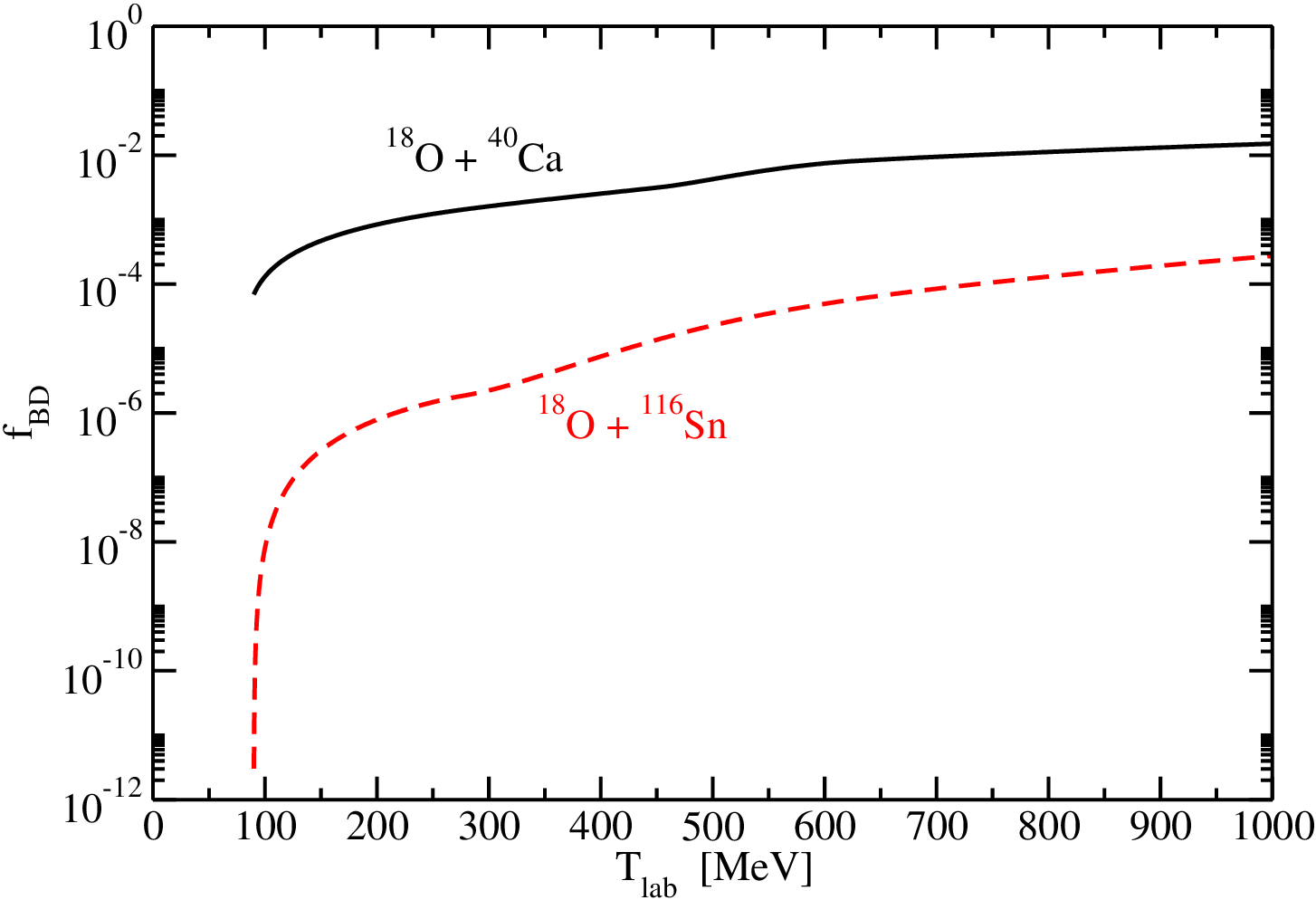}
%fig18_newnew
\caption{(Color online) Left panel: Distortion factor as a function of $R_{abs}$,
 for the same system as in the previous figures.
Right panel: The distortion factor $f_{BD}$ is displayed as a function of
the beam energy for two reaction systems:  $^{18}O+{}^{40}Ca$ and $^{18}O+{}^{116}Sn$. The results were obtained numerically by the ratio of the quantum mechanical DWBA and PWBA forward angle cross sections. }
\end{center}
\end{figure}  
 
%\begin{figure}
%\centering\includegraphics[width=.4\linewidth]{fig19_SCE_fin}
%%fig16_SCE_new1
%\caption{(Color online) The distortion factor $f_{BD}$ is displayed as a function of
%the beam energy for two reaction systems:  $^{18}O+{}^{40}Ca$ and $^{18}O+{}^{116}Sn$. The results were obtained numerically by the ratio of the quantum mechanical DWBA and PWBA forward angle cross sections.
%}
%\end{figure}

Thanks to the factorization of the distortion effects, it is possible to 
isolate the form factor, containing projectile and target transition densities, from the reaction cross section.
It should be noticed that
for L  = 0 transitions, the latter are directly connected to $\beta$ decay strengths. 
 Thus heavy ion SCE reactions are indeed providing access to nuclear matrix elements relevant also for $\beta$-decay. The results derived in this section are of special importance since they are showing explicitly the potential of heavy ion SCE reactions for spectral investigations, including the deduction of nuclear matrix elements for $\beta$-decay.

\section{Results for DCE cross sections}  
Within the hypothesis of a two-step process, calculations have been performed 
for the double charge exchange reaction $^{18}O+{}^{40}Ca\to ^{18}Ne+{}^{40}Ar$ considering just one intermediate state (a 1+ state for both projectile and target).  
Results have been obtained using the FRESCO code \cite{Ian} or by combining the results given by HIDEX for the two single SCE steps. 
The same form factors are employed in the two cases.
Calculations have been performed in PWBA and DWBA approximations, as shown
in Fig.4. 
The two methods (HIDEX double step and FRESCO) are in excellent agreement at the PWBA level.
As far as DWBA calculations are concerned, 
two approximations are employed in the HIDEX case to deal with the distortion effects in the intermediate channel:
i) plane waves are considered;
ii) distorted waves associated with the full (absorptive) optical 
potential are employed.  By comparing with the FRESCO results, one
observes that the ii) option gives the right diffraction pattern, though
the cross section is underestimated by a factor $f_{p.w.}\approx 10$. 
This indicates that distortion effects mainly act only in the entrance and exit channels and should not affect much the virtual intermediate states. 
On the other hand, the option i) reproduces the correct cross section order of magnitude (as given by FRESCO), though the angular pattern is not 
well described.  A quite good agreement with FRESCO is obtained 
by scaling the results of ii) by the factor $f_{p.w.}$. 
The conditions allowing to factorize the DCE cross section described above, 
that would enable to access nuclear matrix elements of similar structure as
(two neutrino) double $\beta$ decay, are presently under 
investigations \cite{CNNP,Elena}. 

\begin{figure}
\centering\includegraphics[width=.5\linewidth]{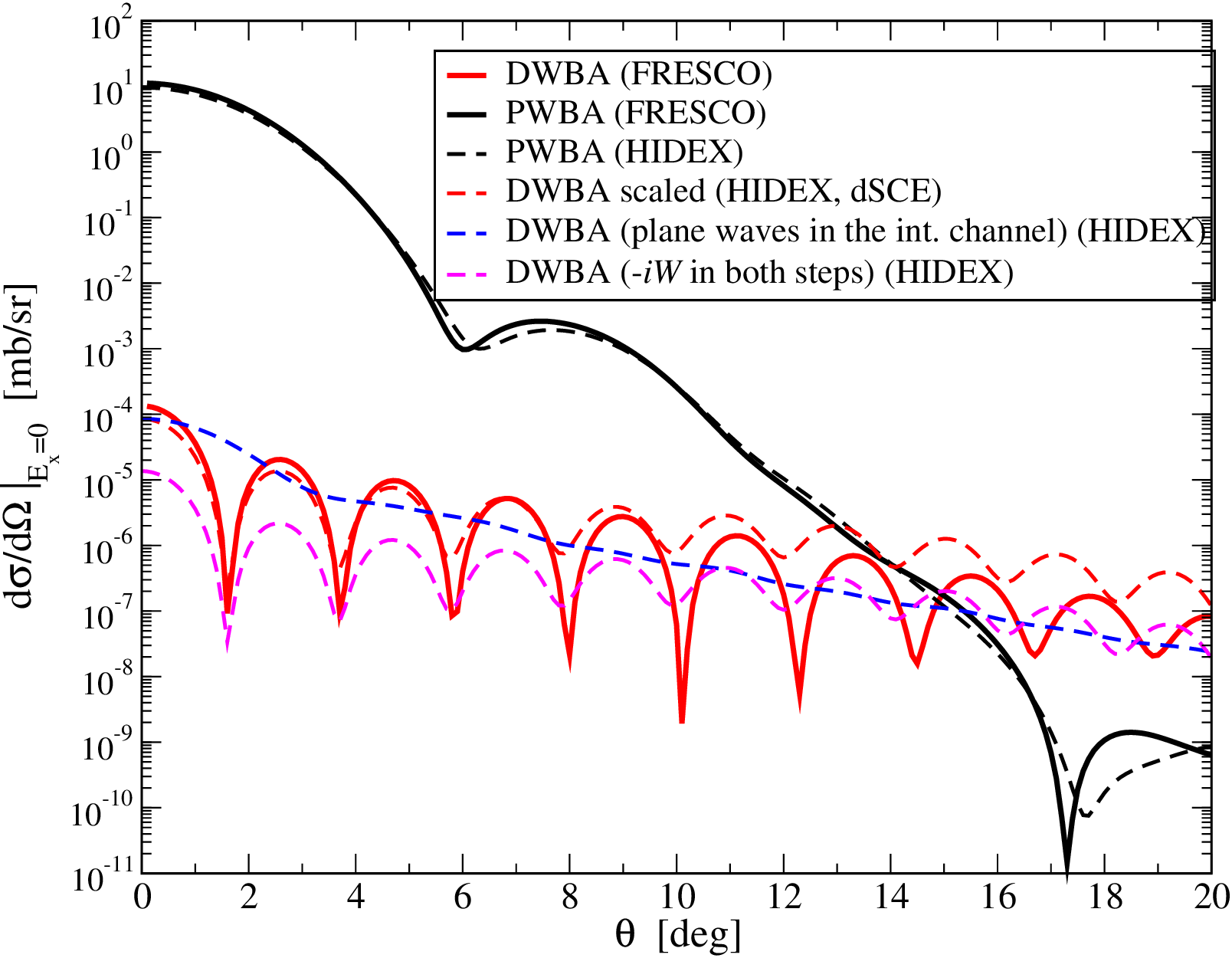}
\caption{(Color online) Angular distribution for the DCE reaction
 $^{18}O+{}^{40}Ca\to ^{18}Ne+{}^{40}Ar$ at 15 AMeV, as obtained with
FRESCO and with double-step Hidex simulations (with two approximations for the
intermediate channel, see text), in PWBA and in DWBA. }
\end{figure}

\section{Conclusions}
In this contribution we discuss new results for heavy ion SCE and DCE reactions. 
As a general feature, heavy ion reactions are characterized by quite large absorption effects, 
justifying the use of the black disk approximation to account for initial and final state
interactions. Then predictions are derived for the behavior of the distortion factor as a function
of beam energy and projectile/target combinations.  

We show that at forward angles the SCE cross section can be factorised, allowing
one to isolate the CE transition matrix element. 
Preliminary results are shown also for DCE reactions, 
depicted as a sequence of two single CE steps. 
The cross section is further reduced for DCE reactions, with respect to SCE processes, as one
can realize by comparing DWBA to PWBA calculations in each case.

% Initial and final state ion-ion interactions are described by optical potentials. Microscopic optical potentials are used, obtained in the impulse approximation, 
%{by folding projectile and target Hartree-Fock-Bogolubov (HFB) ground state densities with free space nucleon-nucleon T-matrices}. 

%Finally, we derive explicit expressions for forward angle cross sections,
%showing that heavy ion SCE reactions are indeed providing access to nuclear matrix elements relevant also for beta-decay. The results derived in this section are of special importance since they are showing explicitly the potential of heavy ion SCE reactions for spectral investigations, including the deduction of nuclear matrix elements for beta-decay.

% QRPA results for nuclear SCE response functions are presented for $^{18}O\to{}^{18}F$ and $^{40}Ca\to{}^{40}K$, respectively.  We discuss results for optical potentials and elastic scattering angular distributions, addressing distortion effects due to ion-ion dynamics, and examples of charge exchange cross sections. As a case of physical interest, we consider throughout the SCE reaction $^{18}O+{}^{40}Ca\to ^{18}F+{}^{40}K$ at $T_{lab}=15$~AMeV. Mathematical-theoretical details have been shifted into a few appendices.

\section*{Acknowledgements}
This project has received funding from the European Union's Horizon 2020 research and innovation programme under grant agreement N. 654002, and from the Spanish Ministerio de Economia y Competitividad and FEDER funds under Project FIS2017-88410-P.


\begin{thebibliography}{99}
\bibitem{Ichimura:2006mq}
  M.~Ichimura, H.~Sakai and T.~Wakasa,
  %``Spin-isospin responses via (p,n) and (n,p) reactions,''
  {\it Prog.\ Part.\ Nucl.\ Phys.}  {\bf 56} (2006) 446.
%  doi:10.1016/j.ppnp.2005.09.001
  %%CITATION = doi:10.1016/j.ppnp.2005.09.001;%%
%%
\bibitem{Thies:2012xg}
  J.~H.~Thies {\it et al.},
  %``High-resolution Mo-100 (He-3, t) Tc-100 charge-exchange experiment and the impact on double-beta decays and neutrino charged-current reactions,''
  {\it Phys.\ Rev.\ C} {\bf 86} (2012) 044309.
%  doi:10.1103/PhysRevC.86.044309
  %%CITATION = doi:10.1103/PhysRevC.86.044309;%%
 %
\bibitem{Frekers:2013yea}
  D.~Frekers, P.~Puppe, J.~H.~Thies and H.~Ejiri,
  %``Gamow-Teller strength extraction from $(^{3}$He, $t)$ reactions,''
  {\it Nucl.\ Phys.\ A} {\bf 916} (2013) 219.
%  doi:10.1016/j.nuclphysa.2013.08.006
  %%CITATION = doi:10.1016/j.nuclphysa.2013.08.006;%%
%
\bibitem{Frekers:2015wga}
  D.~Frekers {\it et al.},
  %``Precision evaluation of the $^{71}$Ga($\nu_e,e^ÃƒÂ¢Ã‹â€ Ã¢â‚¬â„¢$) solar neutrino capture rate from the ($^3$He,$t$) charge-exchange reaction,''
  {\it Phys.\ Rev.\ C} {\bf 91} (2015) 034608.
%  doi:10.1103/PhysRevC.91.034608
%
%\bibitem{Etche:1983}  A. Etchegoyen, D. Sinclair, S. Liu, M.C. Etchegoyen, D.K. Scott and D.L. Hendrie, Nucl. Phys. A \textbf{397} (1983) 343.
%
%\bibitem{Ananta:1986}  N. Anantaraman, J.S. Winfield, S.M. Austin, A. Galonsky, J. van der Plicht, C.C. Chang, G. Ciangaru
%and S. Gales, Phys. Rev. Lett. \textbf{57} (1986) 2375.
%

\bibitem{Fuji:2011}
 Y. Fujita, B. Rubio, and W. Gelletly, {\it Prog. Part. Nucl. Phys.} {\bf 66},
(2011) 549.

\bibitem{Brendel:1988}
  C.~Brendel, P.~von Neumann-Cosel, A.~Richter, G.~Schrieder, H.~Lenske, H.~H.~Wolter, J.~Carter and D.~Sch\"ull,
  %``Quasi-elastic nucleon transfer and single charge exchange in 48 Ti + 42 Ca collisions,''
  {\it Nucl.\ Phys.\ A} {\bf 477} (1988) 162.
%  doi:10.1016/0375-9474(88)90367-3
  %%CITATION = doi:10.1016/0375-9474(88)90367-3;%%
%
\bibitem{Berat:1989amx}
  C.~B\'erat {\it et al.},
  %``Heavy ion charge exchange reactions to probe the giant electric isovector modes in nuclei,''
  {\it Phys.\ Lett.\ B } {\bf 218} (1989) 299.
% doi:10.1016/0370-2693(89)91585-2
  %%CITATION = doi:10.1016/0370-2693(89)91585-2;%
%
\bibitem{Cappuzzello:2015ixp}
  F.~Cappuzzello, M.~Cavallaro, C.~Agodi, M.~Bondi, D.~Carbone, A.~Cunsolo and A.~Foti,
  %``Heavy-ion double charge exchange reactions: A tool toward $0 \nu\beta\beta$ nuclear matrix elements,''
  {\it Eur.\ Phys.\ J.\ A } {\bf 51} (2015) 145.
%  doi:10.1140/epja/i2015-15145-5
%  [arXiv:1511.03858 [nucl-ex]].
  %%CITATION = doi:10.1140/epja/i2015-15145-5;%%
%
%\cite{Cappuzzello:2018wek}
\bibitem{Cappuzzello:2018wek}
  F.~Cappuzzello {\it et al.},
  %``The NUMEN project: NUclear Matrix Elements for Neutrinoless double beta decay,''
  {\it Eur.\ Phys.\ J.\ A} {\bf 54} (2018) 72.
 % doi:10.1140/epja/i2018-12509-3
  %%CITATION = doi:10.1140/epja/i2018-12509-3;%%




\bibitem{Kisa} K. Kisamori {\it et al.}, {\it Phys. Rev. Lett.} 
{\bf 116} (2016) 052501.
\bibitem{Taka} M. Takaki {\it et al.}, CNS Ann. Rep. 94 (2014) 9. 

\bibitem{noi} H. Lenske, J.I. Bellone, M. Colonna, 
J-A. Lay, Theory of Single Charge Exchange Heavy Ion Reactions, 
arXiv:1804.04827, accepted for publication in  {\it Phys. Rev. C}. 

\bibitem{Cappuzzello:2004afa}
  F.~Cappuzzello {\it et al.},
  %``Analysis of the 11B(7Li,7Be)11Be reaction at 57 MeV in a microscopic approach,''
  {\it Nucl. Phys. A} {\bf 739} (2004) 30.
%  doi:10.1016/j.nuclphysa.2004.03.221

\bibitem{Ian} I. Thompson,
{\it Comp. Phys. Rep.} {\bf 7} (1988) 167.


\bibitem{CNNP} H. Lenske, 
{\it J. Phys. Conf. Ser.}
{\bf 1056} (2018) 012030; 
 J.I. Bellone, M. Colonna, H. lenske,  J-A. Lay,
{\it J. Phys. Conf. Ser.} 
{\bf 1056} (2018) 012004.

 \bibitem{Elena} E. Santopinto, H. Garc\'ia-Tecocoatzi, R. I. Magana-Vsevolodovna, J. Ferretti, 
Heavy-ion double-charge-exchange and its relation to neutrinoless double-beta decay, arXiv:1806.03069.


%\bibitem{bib:voss2005} Herbert Vo{\ss},
%  \emph{Math mode}, available at the URL
%\url{http://www.tex.ac.uk/tex-archive/info/math/voss/mathmode/Mathmode.pdf}.

%\bibitem{bib:mittelbach2004} Frank Mittelbach and Michel Goossens,
%  \emph{The \LaTeX{} Companion}, 2nd ed. (Addison-Wesley, Boston,
%  2004).
%\bibitem{bib:pakin2003} Scatt Pakin,
%  \emph{The Comprehensive \LaTeX{} Symbol List}, available at the URL
%\url{http://www.ctan.org/tex-archive/info/symbols/comprehensive/symbols-a4.pdf}.
\end{thebibliography}
\end{document}